\documentclass[preprint,nofootinbib,aps,superscriptaddress,eqsecnum]{revtex4-1}
\pdfoutput=1
\usepackage[colorlinks=true,citecolor=blue,linkcolor=blue,breaklinks=true]{hyperref}
\usepackage{graphicx}
\usepackage{mathrsfs}
\usepackage{soul}
\usepackage{natbib}
\usepackage{enumerate}
\usepackage{amsmath,amssymb}
\usepackage{slashed}
\usepackage{amssymb,amsmath,subfigure,xcolor}

\newcommand{\lapprox}{%
\mathrel{%
\setbox0=\hbox{$<$}
\raise0.6ex\copy0\kern-\wd0
\lower0.65ex\hbox{$\sim$}
}}
\newcommand{\gapprox}{%
\mathrel{%
\setbox0=\hbox{$>$}
\raise0.6ex\copy0\kern-\wd0
\lower0.65ex\hbox{$\sim$}
}}
\newcommand{\ba}{\begin{array}}
\newcommand{\ea}{\end{array}}
\newcommand{\bd}{\begin{displaymath}}
\newcommand{\ed}{\end{displaymath}}
\newcommand{\beq}{\begin{equation}}
\newcommand{\eeq}{\end{equation}}
\newcommand{\bea}{\begin{eqnarray}}
\newcommand{\eea}{\end{eqnarray}}

\newcommand{\ra}{\rightarrow}
\newcommand{\nn}{\nonumber}

\def\a{\alpha}

\def\g{\gamma}

\def\m{\mu}
\def\n{\nu}

\newcommand{\mdm}{m_{\text{DM}}}
\newcommand{\mbh}{M_{\text{BH}}}
\newcommand{\msol}{M_{\text{sol}}}
\newcommand{\mso}{{\rm M}_{\odot}}
\newcommand{\mh}{M_{\text{halo}}}
\newcommand{\gfp}{G_F^\prime}

\def\q2 {q^2}

\def\bt{\begin{table}}
\def\et{\end{table}}

\catcode`@=11 
\def \gsim{\mathrel{\mathpalette\@versim>}}
\def \lsim{\mathrel{\mathpalette\@versim<}}
\def \@versim#1#2{\lower0.4ex\vbox{\baselineskip\z@skip\lineskip\z@skip
     \lineskiplimit\z@\ialign{$\m@th#1\hfil##\hfil$%
     \crcr#2\crcr\sim\crcr}}}
\catcode`@=12 

\begin{document}

\renewcommand*{\thefootnote}{\fnsymbol{footnote}}

\begin{center}

{\large\bf Astronomy with energy dependent flavour ratios of extragalactic neutrinos}
\\[15mm] 
Siddhartha Karmakar\footnote{E-mail:  karmakars@iitb.ac.in},
Sujata Pandey\footnote{E-mail:  phd1501151007@iiti.ac.in} and Subhendu
Rakshit\footnote{E-mail: rakshit@iiti.ac.in} 
\\[2mm]

{\em Discipline of Physics, Indian Institute of Technology Indore,\\
 Khandwa Road, Simrol, Indore - 453\,552, India}
\\[20mm]
\end{center}

\begin{abstract} 
\vskip 20pt

High energy astrophysical neutrinos interacting with ultralight dark matter (DM)  can undergo flavour oscillations that induce an energy dependence in the flavour ratios. Such a dependence on the neutrino energy will reflect in the track to shower ratio in neutrino telescopes like IceCube or KM3NeT. This opens up a possibility to study DM density profiles of  astrophysical objects like AGN, GRB etc., which are the suspected sources of such neutrinos.

\end{abstract}
 \vskip 1 true cm
 \pacs {}
\maketitle
\section{Introduction}
\label{intro}
Traditional astronomy based on photons ceases to work for very high energy gamma rays, above a few tens of TeV, as they get absorbed interacting with background photons on their way to the Earth. Hence, it is rather difficult to gather first hand information about the interiors of the astrophysical objects like active galactic nuclei (AGN), gamma ray bursts (GRB), etc. at very high energies. However, these objects are expected not only to emit photons, but also cosmic rays and neutrinos with extreme energies stretching up to EeVs or more. As neutrinos interact only weakly, astronomy with high energy astrophysical neutrinos seem quite promising. IceCube has seen such neutrinos up to a few PeV and future upgrades are designed to improve the statistics~\cite{Aartsen:2016xlq,Aartsen:2014gkd,Aartsen:2019mbc,Aartsen:2020fgd}.  We have already been able to `look' into the interiors of the Sun and the supernova 1987A through neutrinos and now we aspire to do the same for these astrophysical objects. However, the matter density in these astrophysical objects is usually too low to affect neutrino propagation~\cite{Lunardini:2000swa}. While this allows the neutrinos to stream out of these objects unhindered,  little information about the interiors are usually carried by these neutrinos. This is the main stumbling block of astronomy with the high energy astrophysical neutrinos, compared to the same with photons. We propose in this paper that if the dark matter is ultralight, then even a feeble interaction of neutrinos with DM inside these objects may help circumvent such shortcomings of neutrino astronomy. 

In the standard scenario, astrophysical neutrinos are produced from charged pion decay, yielding a flavour ratio $\nu_e:\nu_\mu:\nu_\tau=1:2:0$. Then these neutrinos undergo vacuum flavour oscillations to reach earth with a flavour ratio $\nu_e:\nu_\mu:\nu_\tau=1:1:1$, independent of the energy of the neutrinos. We show that this picture takes a blow once these neutrinos are allowed to interact with a surrounding ultralight dark matter halo. 

Building models for neutrino-dark matter interactions that lead to appreciable flux suppression is rather challenging~\cite{Pandey:2018wvh}. Such interactions can lead to a  lack of temporal coincidence between the observation of electromagnetic signals and neutrinos from GRBs~\cite{Koren:2019wwi,Murase:2019xqi}. On the other hand, the strength of such interactions may be feeble enough to lead to any appreciable flux suppression at the IceCube, but these can severely affect neutrino oscillations in regions where the dark matter number density is significant. We show that this leads to an energy dependent flavour ratio which drastically differs from the standard expectation of flavour-universal flux of such neutrinos. This also predicts different flavour ratios for neutrinos and anti-neutrinos. Although it is the dark matter interactions with the neutrinos that influence the neutrino oscillations, to match the standard literature, we refer to this as `matter effects' in this paper.

The role of DM-neutrino interactions to preserve the source flavour ratio during propagation has been reported~\cite{Farzan:2018pnk}.
The fact that such interactions might help in finding out DM distribution is not that surprising. But the fact that it does so by imprinting the dark matter halo profile in the energy dependence of neutrino flavour ratios is rather intriguing. 
Alternatively, such an energy dependence  might even  originate at the source~\cite{Mehta:2011qb, Hummer:2010ai}
 Various implications of the measurement of  neutrino flavour ratios at IceCube have been studied in the literature, such as constraining certain new physics scenarios~\cite{Bustamante:2015waa, Brdar:2018tce, Miranda:2013wla, Arguelles:2015dca}, neutrino decay~\cite{Pagliaroli:2015rca,Bustamante:2016ciw,Sadhukhan:2018nsk,Denton:2018aml,Abdullahi:2020rge}, testing the unitarity of mixing matrix~\cite{Ahlers:2018yom}, contributions from exotic sources~\cite{Mena:2006eq,Carpio:2020app}, various neutrino-DM interactions~\cite{Farzan:2018pnk, deSalas:2016svi,  Rasmussen:2017ert}, etc.

The paper is organised as follows: In the next section, we discuss certain general aspects of the neutrino sources relevant for this paper, such as DM density profile, relation to black hole mass, etc. In Sec.~\ref{sec:III} we lay down the formalism to evaluate the neutrino flavour ratios in the presence of a DM potential. The energy dependencies of flavour ratios and track to shower ratio at IceCube have been discussed in Sec.~\ref{sec:IV}. Subsequently, we summarise our key findings and eventually conclude.

\section{Neutrino Astronomy}
\label{sec:II}

Weakly interacting massive particles, posing as cold dark matter\,(CDM) candidates, seem to fit cosmological observations quite well, though their searches at direct and indirect dark matter experiments and at colliders have not lead to any success so far. Moreover, on smaller galactic scales, the prediction of the CDM models seem
to disagree with the observations~\cite{Salucci:2002nc,Blok:2002tr,Tasitsiomi:2002hi,Navarro:1995iw,deBlok:2009sp}. Namely, the CDMs predict a cuspy profile at the centres of galaxies in which the DM density  $\rho$ is expected to fall with the radius $r$ as $1/r$, whereas the observations suggest $\rho\sim r^0$, indicating the existence of a `core'~\cite{Navarro:1995iw,deBlok:2009sp}. Moreover, the predicted luminosities of the biggest satellite galaxies are significantly higher than their observed values~\cite{BoylanKolchin:2011de}.

Ultralight scalar DM is an attractive candidate to address these small-scale problems~\cite{Harko:2011xw,Boehmer:2007um,Matos:2000ss,Guzman:2013coa}. 
The observations of Lyman-$\a$ forest~\cite{Irsic:2017yje}, CMB spectrum~\cite{Li:2013nal}, and the supermassive black hole~(SMBH) M87$^*$~\cite{Davoudiasl:2019nlo} exclude DM masses lower than $\mdm \sim 10^{-22}$~eV.  
Ultralight scalar DM of masses $\mdm \gtrsim 10^{-22}$~eV are viable in the presence of DM self-interactions~\cite{Arbey:2003sj,Peebles:2000yy,Goodman:2000tg,Fan:2016rda,Harko:2011xw,Lora:2011yc}. 
Although at cosmological scales ultralight scalar DM behave as CDM,  at smaller scales, depending on  their de-Brogli wavelengths, they behave differently. 
These particles can exist in the form of a Bose-Einstein condensate, the associated quantum pressure thereby compensating the gravitational pressure, forming a core-like structure at the centre~\cite{Hu:2000ke,Peebles:2000yy,Matos:2007zza,Su:2010bj,Matos:2007zza}. This core has an uniform density at the centre, but the density falls abruptly at some radial distance. This has been confirmed by numerical simulations. 
 Various analytical calculations of ultralight DM profile, both in the presence and the absence of a BH at the centre of the galaxy has been carried out~\cite{Davies:2019wgi,Schive:2014hza,Bar:2019pnz}. While presenting our results we have considered following DM profile of the solitonic core, in the presence of a SMBH of mass $\mbh$~\cite{Davies:2019wgi}:
\beq
\rho(r)=\rho_0 \exp(-r/a), \label{rho}
\eeq
where 
\beq
a=\frac{1}{G \mbh \mdm^{2}} \label{a}
\eeq
and $\rho_{0}$ is related to the mass of the solitonic core $\msol$ as
\beq
\rho_{0}=\frac{\msol  }{8\pi a^3} \,\,. \label{rho_definition}
\eeq
 Note that, the DM density profile in eq.~\eqref{rho} is valid when the mass of SMBH is larger than the soliton mass.

The empirical relationship between $\mbh$ and $\mh$ is derived from the correlation of $\mbh$ with Sersic index and stellar velocity dispersion~\cite{Bandara:2009sd}. 
The empirical $\mbh-\mh$ relation from ref.~\cite{Bandara:2009sd} reads
\bea
\log (\mbh/\mso)= \alpha + \beta \Big[\log (\mh/\mso)-13 \Big]
\eea
where $\alpha = 8.17 \pm 0.4$ and $\beta= 1.57 \pm 1.2$ at 3$\sigma$ CL. Slightly refined versions of such a relationship also exist in the literature~\cite{Kormendy:2013dxa,Sun:2013hvd}. 
It has been argued that the mass of solitonic core scales with DM halo mass as $\msol \propto \mh^{1/3}$~\cite{Schive:2014hza,Bar:2019pnz}. 
There have also been attempts to predict the shape of the core using velocity dispersion~\cite{Davies:2019wgi,Bar:2019pnz}. Though we do not use these scaling relations while choosing the benchmark values of $a$ and $\rho_0$ due to the caveats discussed later in this section. 

 We show that the future neutrino telescopes can complement the traditional telescopes providing valuable inputs related to the shape of the DM profiles of various astrophysical objects. The core models of AGN are examples of the kind of astrophysical objects we are referring to. Although, the acceleration mechanism for the cosmic rays and sites for shocks are not known, for our purpose, it is safe to consider a situation in which neutrinos are produced from the charged pions originating from the interaction of accelerated protons with photons in the corona around a distance $\sim$ 10--40$R_s$, where $R_s=2G\mbh$ is the Schwarzschild radius of the black hole~\cite{Inoue:2019fil,Kimura:2019yjo,Kimura:2020thg}. 

To get a feel for the length scales involved here, let us consider $\mbh\sim 10^5 \mso$, corresponding to $R_s\sim 5\times 10^{-8}$~pc. Neutrino emission takes place around a distance $10^{-7}$~pc from the centre, where the DM density of the solitonic core is uniform, considering $\mdm\sim 3\times 10^{-17}$~eV. According to eq.~\eqref{a},  this combination of $\mbh$ and $\mdm$ leads to $a=10^{-6}$~pc, around where the core meets its edge. After this radial distance, the density of DM halo is drastically less. 
The sharp fall in the DM density at the edge can induce non-adiabaticity in the neutrino oscillation probability, which will in principle make neutrino astronomy possible determining the shape of the core. Here, the oscillation length of a neutrino of energy 1~PeV can only be as large as $\sim 10^{-12}$~pc. This reaffirms the fact that the oscillations do get averaged out while these neutrinos come out of these astrophysical objects.

As mentioned earlier, due to the low matter density in these environments, the standard matter effect due to electrons is negligible~\cite{Lunardini:2000swa}. But the ultralight mass of the DM results in a sizable number density, leading to a substantial matter effect from $\nu$--DM interactions. As we will show in the next section, the effective mass term induced by such interactions drastically affect neutrino oscillations, so that the DM profile gets imprinted on the energy dependence of the flavour ratios of neutrinos detected at neutrino telescopes.

The aforementioned relations between the SMBH, halo, and soliton masses are under substantial investigation in the literature. We would like to point that,  while evaluating the core-halo relation ref.~\cite{Schive:2014hza} consider non-interacting BEC DM which forms solitonic core at the galactic centre in the absence of a BH. Therefore,  this relation holds only in the case where solitonic cores are formed before SMBH~\cite{Davies:2019wgi, Chavanis:2019amr, Bar:2019pnz}. Further, the $\msol$-$\mh$ is valid for a very small range of $\mdm$, around $10^{-22}$~eV~\cite{Bar:2019pnz}. As an example, for $\mdm \sim 10^{-19}$~eV, the value of soliton mass can be smaller compared to $\msol$ predicted by ref.~\cite{Schive:2014hza} by an order of magnitude, even in the absence of SMBH~\cite{Bar:2019pnz}. Therefore, in this paper we consider $\rho_0$ and $a$ as independent parameters to describe DM density profile at source as long as $\mbh > \msol$.

\section{Neutrino Oscillations in a dark matter halo}
\label{sec:III}
Various aspects of neutrino-DM interactions have been studied in the literature~\cite{Pandey:2018wvh,  
Koren:2019wwi,Farzan:2018pnk,Berlin:2016woy,Nieves:2019izk,Murase:2019xqi,Blennow:2019fhy,Alvey:2019jzx,Choi:2019ixb,Arguelles:2019ouk,Cherry:2014xra,Chauhan:2018dkd}. An encyclopedia of  interactions of neutrinos with ultralight scalar DM leading to an effective vertex  $\nu$-$\bar{\nu}$-$\phi$-$\phi^{*}$ can be found in ref.~\cite{Pandey:2018wvh}. 
Most of these interactions are severely restricted by the ensuing interactions of the corresponding charged leptons implied by the $SU(2)_L$ invariance.

As an example, for the vectorial type of neutrino-DM interaction, the constraints on the effective strength  $\epsilon$ are as follows: In order to avoid  anomalous energy loss in sun, one must ensure $\epsilon_{ee} \lesssim 10^{-38}$~eV$^{-2}$~\cite{Harnik:2012ni}.
LHC bounds from heavy $Z'$ searches can be used to obtain $\epsilon_{\mu \mu} \lesssim 1.5 \times 10^{-26}$~eV$^{-2}$~\cite{Sirunyan:2018nnz}. Bounds on flavour violating charged lepton decays translate to $\epsilon_{\mu \tau} \lesssim  10^{-31}$~eV$^{-2}$~\cite{Heeck:2016xkh}, $\epsilon_{\mu e} \lesssim 10^{-40}$~eV$^{-2}$~\cite{Farzan:2015hkd},  and $\epsilon_{\tau e} \lesssim 4 \times 10^{-32}$~eV$^{-2}$~\cite{Farzan:2015hkd}.
 On the other hand, the constraints on $\epsilon_{\tau \tau}$ are comparatively less stringent. The most stringent bound on $\epsilon_{\tau \tau}$ comes from the measurement of partial $Z$ decay width $\Gamma(Z \ra \tau^{+} \tau^{-})$ which reads $\epsilon_{\tau \tau} \leqslant 1.3 \times 10^{-20}$~eV$^{-2}$~\cite{Pandey:2018wvh}. 
Hence, we explore the possible impact of  matter effect     on the flavour ratio of astrophysical neutrinos due to $\epsilon_{\tau \tau}$, at IceCube and future neutrino observatories.  
For very light dark matter, depending on the model behind neutrino--DM interactions, given the rather relaxed limit mentioned above, $\epsilon_{\mu \mu}$ can also influence neutrino oscillations. Although for simplicity, here we will consider only $\epsilon_{\tau \tau}$ to be non-zero, the analysis can easily be extended to incorporate effects of $\epsilon_{\mu \mu}$ as well. Note that, to prevent ultralight DM from being thermalised in the primordial soup, the Big Bang nucleosynthesis~(BBN) constraint demands $\epsilon \lsim 6 \times 10^{-22}$~eV$^{-2}$ for all flavours. 

In passing, a comment on feasibility to build such a model that allows only the third generation leptons to interact with DM seems quite pertinent. Such a scenario can easily be envisaged if such an interaction is mediated by a $Z^\prime$ vector boson, that pertains to a gauged $U(1)_\tau$ symmetry~\cite{Pandey:2018wvh}:
\beq
\mathcal{L} \supset  ig' (\phi^{*} \partial_{\mu} \phi-\phi \, \partial_{\mu} \phi^{*}) Z^{\prime \m} +  f \bar{\nu}_{\tau} \g_{\m} \n_{\tau} Z^{\prime \m} \, .
\label{L}
\eeq
This leads to an interaction strength $\gfp= g^\prime f/m_{Z^\prime}^2$, for $\sqrt{s}\ll m_{Z^\prime}$. 
Henceforth, $G'_F$ is synonymous with $\epsilon_{\tau\tau}$ in this paper.
While passing through the DM halo,  neutrinos will experience a potential 
\beq
V_{\tau\tau}= \frac{\gfp}{\mdm} \rho(r) \, .
\label{eq:Vtt}
\eeq
Several other models can also lead to $\nu$-DM interactions of desired strength~\cite{Pandey:2018wvh}.
For these other interactions with different momentum dependencies, the resultant potentials simply differ by factors of $\mdm$.  
 
In the presence of potential $V_{\tau\tau}$, the Hamiltonian governing the evolution of neutrinos is augmented by a `matter' term as follows:
\bea
H_{\rm eff} = \frac{1}{2 E (1+z)}  U \begin{pmatrix}
0 & 0 & 0\\
0 & \Delta m_{12}^2 & 0  \\
0 & 0 & \Delta m_{32}^2 \\
\end{pmatrix} U^{\dagger} 
- 
\begin{pmatrix}
0 & 0 & 0\\
0 & 0 & 0  \\
0 & 0 & V_{\tau \tau} (r)  \\
\end{pmatrix} ,
\label{Heff}
\eea
where $U$ stands for the PMNS matrix in vacuum and the redshift $z$ is indicative of the location of the neutrino source from the Earth. $E$ is the energy of the neutrino at earth. 
Note that for antineutrinos, $V_{\tau\tau}$ flips its sign.

$H_{\rm eff}$ has to be diagonalised to compute the modified PMNS matrix in the presence of the DM potential. Owing to the extreme density at the core of these astrophysical objects, the first term in eq.~\eqref{Heff} is negligible compared to the second term for extremely energetic neutrinos. As a result, $U^S$, the PMNS matrix at the source of production, becomes an identity matrix at substantially high energies. At the Earth, however, both the terms in eq.~\eqref{Heff} should be taken into account for computation of $U^D$, the PMNS matrix at the detector. In this case, $z=0$, and $U^D$ depends on energy of the neutrino as recorded on earth and  the combination ${\gfp}/{\mdm}$, that decide the value of $V_{\tau \tau}$ at the detector. 

Large DM density at the core  leads to a considerable shift of the values of the effective mixing angles from the vacuum mixing angles. Moreover, a sharp change in the density profile may give rise to non-adiabaticity. Neutrino flavour oscillation probability is given by,
\bea
P_{\alpha \beta}=   |U^{D}_{\beta i}|^2 |U^{S}_{\alpha i}|^2 &-& P^{c}_{ij} (|U^{D}_{\beta i}|^2 - |U^{D}_{\beta j}|^2) (|U^{S}_{\alpha i}|^2 - |U^{S}_{\alpha j}|^2)  \nn\\
&-& P^{c}_{ik} P^{c}_{kj} (|U^{D}_{\beta k}|^2 - |U^{D}_{\beta j}|^2) (|U^{S}_{\alpha i}|^2
- |U^{S}_{\alpha k}|^2) \,.
\label{prob}
\eea
Here, we use the convention of summation over repeated indices, which will be followed in the rest of the paper too. Note that, 
The first term on the right  is the adiabatic contribution after neutrino oscillations get averaged out as the oscillation length is much less than the distance traversed. The rest of the terms contribute only when some non-adiabaticity is present. $P^{c}_{ij}$ stands for the jumping probability between the two mass eigenstates $\n_i$ and $\n_j$ and is given by~\cite{Kaneko:1987zza, Toshev:1987jw, Kuo:1989qe, Notzold:1987cq}:  
\bea
P^{c}_{ij}= \frac{\exp(- \frac{\pi}{2} \gamma_{ij}^{R} F_{ij})- \exp(-\frac{\pi}{2} \gamma_{ij}^{R} \frac{F_{ij}} {\sin^2 \theta_{ij}})}{1- \exp(-\frac{\pi}{2} \gamma_{ij}^{R} \frac{F_{ij}}{\sin^{2} \theta_{ij}}) },
\label{pc}
\eea
where $\gamma_{ij}$ is the non-adiabaticity parameter, which at the resonance is given by
 \beq
 \gamma_{ij}^{R} = \frac{\Delta m^2_{ij} \sin^{2} 2\theta_{ij}}{2 E (1+z) \cos 2\theta_{ij}\, |\text{d} \ln \rho/\text{d} r|_{R}}\, .
 \label{adr}
 \eeq
$\gamma_{ij}\sim 0$ corresponds to extreme non-adiabaticity. 
A significant amount of non-adiabaticity, leading to transitions between different mass eigenstates, can  be induced in our case due to the interplay of the extreme energy of the neutrinos and the density gradient of DM at the edge of the solitonic core inside an AGN. For the profile given by eq.~\eqref{rho}, $|\text{d} \ln \rho/\text{d} r|_R=1/a$ and~\cite{Kuo:1989qe,Kuo:1988pn}    
\beq
F_{ij}= \frac{4}{\pi}\, \text{Im} \int_{0}^{i} db \frac{(b^2+1)^{1/2}}{(b \tan 2 \theta_{ij} +1)}=\begin{cases}1- \tan^{2}\theta_{ij}, &  ~\text{if}  ~~\theta_{ij} \leqslant \pi/4\\
1- \cot^{2}\theta_{ij}, & ~\text{if} ~~\theta_{ij}> \pi/4\,\,\,.
\end{cases}
\eeq

Energy dependence in $P_{\alpha \beta}$ creeps in through  $U^{D}$ and $P^{c}_{ij}$.
Note that in eq.~\eqref{prob},  $P_{\alpha \beta}$ denotes the probability of oscillation from the flavour $\alpha$ to $\beta$  and it differs from $P_{\beta\alpha}$ due to the different DM densities at the source and the detector.
  The jumping probability is negligible between the states that never undergo resonance even in the presence of matter potential. 
One needs to be cautious in reading off eq.~\eqref{prob}. Here,  for the non-adiabatic contributions only those terms relevant for the scenario has to be taken into account. For example, if the $31$ resonance is followed by a $32$ resonance, only the term $P^{c}_{32} P^{c}_{31}$ has to be included in the term responsible for two resonances. In principle, depending on the density profile of the DM, eq.~\eqref{prob} can easily be extended to include terms with more than two resonances.

While presenting numerical estimates we use the following set of parameters obtained from a global fit~\cite{nufit4.1, Esteban:2018azc} of solar, atmospheric, reactor and accelerator neutrino oscillation experiments~: $\theta_{12}=33.8^\circ$, $\theta_{23}=48.6^\circ$, $\theta_{13}=8.6^\circ$, $\delta_{\rm CP}=1.22\pi$~rad.  This set corresponds to the normal hierarchy of the neutrino masses, with $ \Delta m_{32}^2 =  m_3^2-m_2^2=2.53\times 10^{-3}$~eV$^2$ and $\Delta m_{21}^2 = m_2^2-m_1^2=7.39\times 10^{-5}$~eV$^2$.

The galactic DM density changes very slowly with respect to position, leading to negligible jumping probability near the detector. For galactic scales the ultralight DM behaves similar to CDM~\cite{Ferreira:2020fam, Magana:2012ph}. The  isothermal dark matter profile is~\cite{Bahcall:1980fb}
$$\rho_\text{Iso}= 0.4 ~\Big( \frac{1+(8.5/2)^2}{1+(r/2)^2} \Big)~\rm{GeV cm^{-3}},$$
 where $r$ is the radial distance from the Galactic Center in kpc. Thus, the gradient of DM density is $|\text{d} \ln \rho_\text{Iso}/\text{d} r| = \frac{1}{2} \frac{r}{(1+(r/2)^{2})}~\rm{kpc^{-1}}$  and the maximum value of $|\text{d} \ln \rho_\text{Iso}/\text{d} r|$ is around $ 0.5$~kpc$^{-1}$, which leads to adiabatic propagation of neutrinos within the galaxy for the entire energy range considered in the paper. Hence, the flavour ratio is independent of local DM density profile and depends only on the density of DM at Earth $\rho_D \simeq 0.4$~GeV cm$^{-3}$~\cite{Read:2014qva}.  Hence, for the DM profile considered here, there are four quantities that determine the probability of oscillations: the redshift $z$, the parameters related to the solitonic core $\rho_0$ and $a$, and the combination ${\gfp}/{\mdm}$.

\section{Energy dependence of flavour ratios}
\label{sec:IV}
Neutrino flavour ratios at the detector is related to the same at source as follows
\beq
f_\beta^D = P_{\alpha\beta} f_\alpha^S. \label{fratio}
\eeq
Clearly, in the case of vacuum oscillations, for the source flux ratio $1:1:1$, the flavour ratio at earth remains $1:1:1$. If the ratio at source is $1:2:0$, then the flavour ratio at the detector
\beq
f_\beta^D =  P_{e \beta}+ 2 P_{\mu \beta}
= |U_{\beta i}|^2 (|U_{ei}|^2 + 2 |U_{\mu i}|^2) .  
\eeq
This leads to $f_e^D:f_\mu^D:f_\tau^D\simeq 1:1:1$.  For $\theta_{13}=0$ and maximal $\nu_2-\nu_3$ mixing, the equality is exact. While these favour ratios at earth are energy independent, as we will further discuss, neutrino-DM interactions may induce an energy dependence.

In the presence of large matter effect,  the mixing matrix at the source deviates significantly  from that at vacuum. 
 In matter,
 \bea
  \sin 2\theta_{13}^{M} = \Delta m_{31}^2 \sin 2\theta_{13}/[(2E(1+z) V_{\tau \tau} - \Delta m_{31}^2 \cos 2 \theta_{13})^2+ (\Delta m_{31}^2 \sin 2\theta_{13})^2]^{1/2}
  \label{eq:mattermixing}
  \eea
   and as $E$ is large,  $2E (1+z) V_{\tau \tau} \gg \Delta m_{31}^2 \cos 2 \theta_{13}$, leading to a vanishingly small $\sin 2\theta_{13}^{M}$.    Similarly, $\sin 2\theta_{23}^{M}$ also becomes small at large values of $E$.
As $E$ increases further, the vacuum oscillation term of the Hamiltonian can be neglected and the mixing matrix tends to identity. 

In the case of adiabatic oscillation, due to a large matter effect induced by neutrino-DM interaction, the flavour ratios of the (anti)neutrinos at the source are preserved~\cite{Farzan:2018pnk}. Here we focus on a more general and interesting possibility of non-adiabatic flavour transition which can change flavour ratio at IceCube. 
As eq.~\eqref{prob} indicates, the probability of flavour transition in such a general scenario depends on $U^S$, $U^D$, $P^{c}_{31}$, and $P^{c}_{32}$.  These in turn depends on   $G_{F}^{\prime}/\mdm$, $a$, and the DM density at the detector and source. 
As discussed in Sec.~\ref{sec:III}, with the best-fit values of the PMNS mixing angles considered here, $\theta_{23}$ and $\theta_{13}$ lie in the second and first octant respectively.
 Thus, the resonance condition
\bea 
  2 E^R_{ij} (1+z) V_{\tau \tau}=  \Delta m_{ij}^2 \cos 2\theta_{ij}
  \label{eq:resonance}
\eea  
  is satisfied for the negative and positive values of the potential for $ij = 32$ and $ij = 31$ respectively. 
 According to eq.~\eqref{adr}, for a fixed value of $E$, the condition for non-adiabatic oscillation, $\gamma^{ij}_R \lesssim 1$ is satisfied for two different values of $a$ for the 32 and 31 transitions.  
   At $E \sim 1$~PeV, for $V_{\tau \tau} > 0$, $\gamma_{31} \sim 1$ is obtained for $a \sim 10^{-3}$~pc, whereas, for $V_{\tau \tau} < 0$, $\gamma_{32} \sim 1$ for $a \sim 10^{-5}$~pc. Non-adiabaticity for a lower energy can be obtained by lowering the value of $a$.
 Henceforth, we consider $V_{\tau \tau}$ to be positive for neutrinos and thus, negative for antineutrinos. In the following, we discuss the energy dependence of the flavour ratios for different values of $a$ with $\rho_0 = 7.4 \times 10^{-3}$~eV$^4$ and $G_{F}'/ \mdm = 10^{-13}$~eV$^{-3}$.

\begin{figure}[h!]
 \begin{center}
\subfigure[]{
 \includegraphics[width=2.6in,height=2.0in, angle=0]{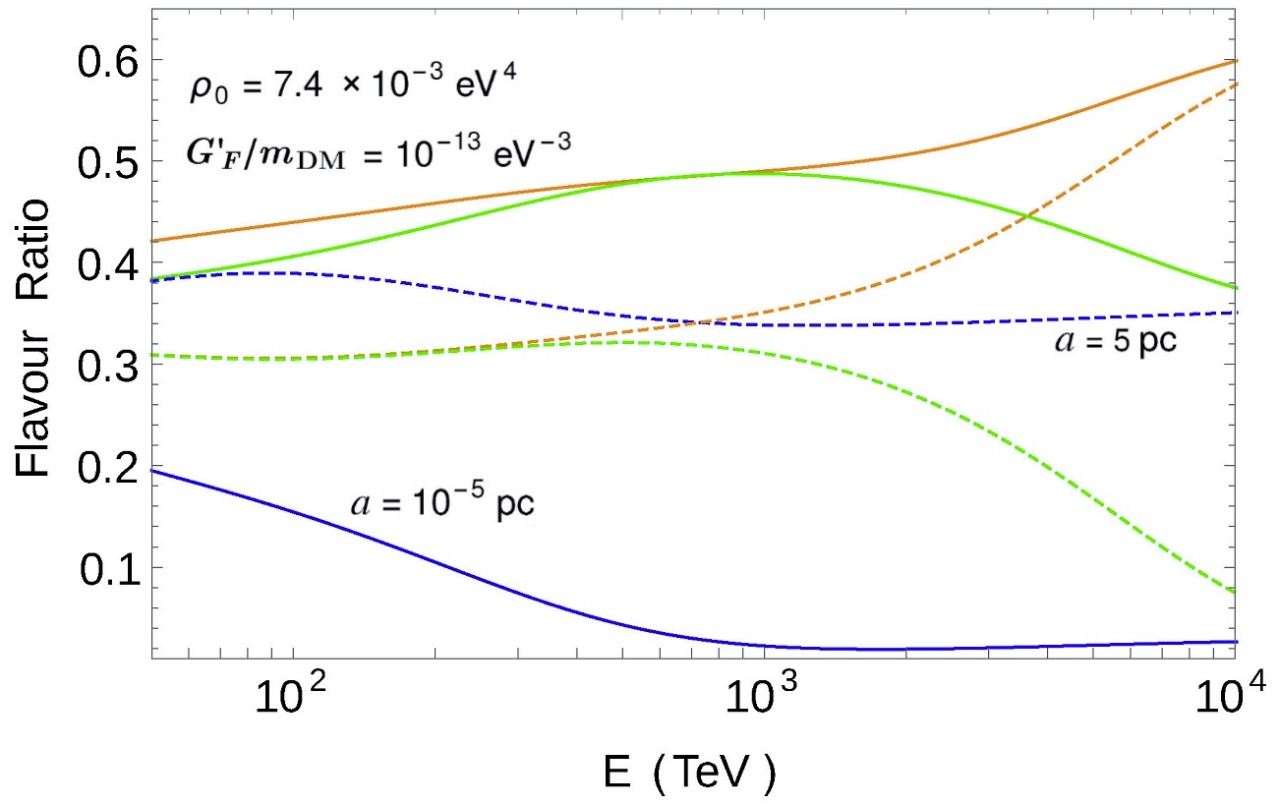}}
 \hskip 15pt
 \subfigure[]{
 \includegraphics[width=2.6in,height=2.0in, angle=0]{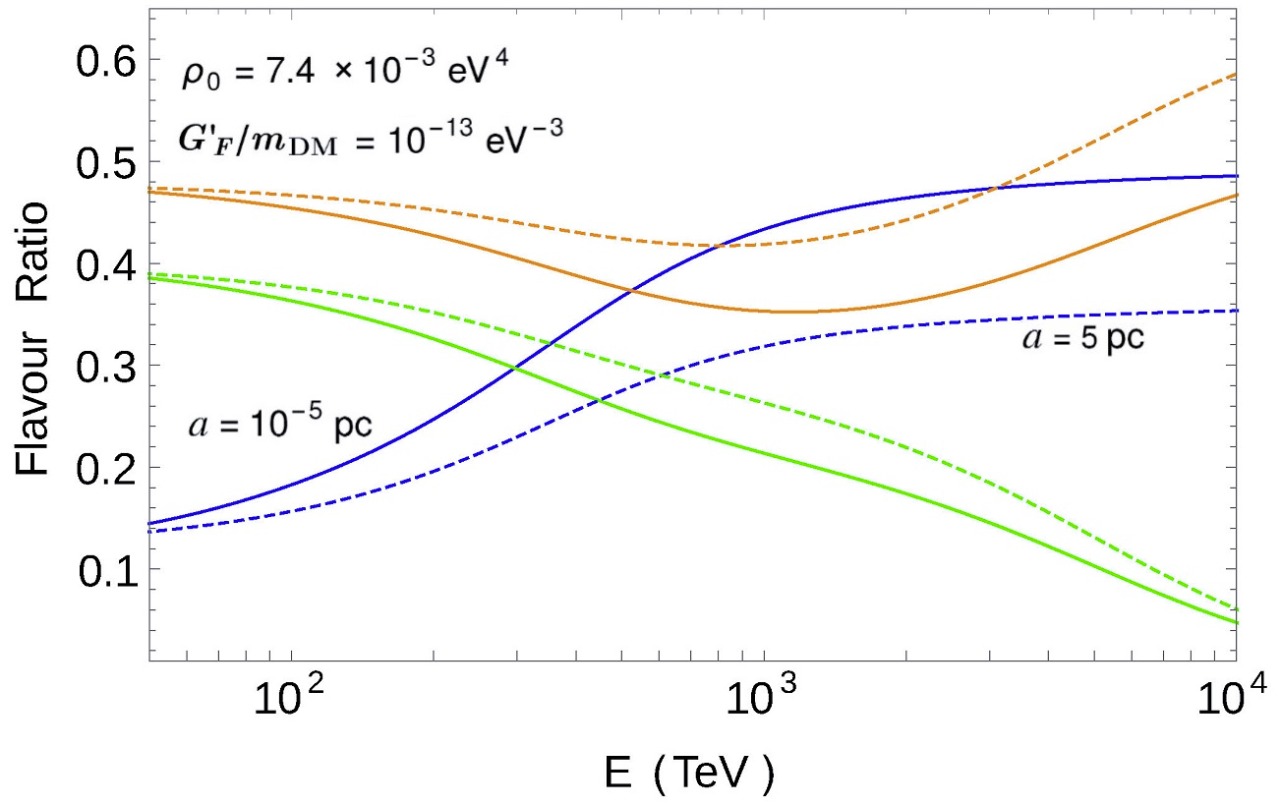}}
 \subfigure[]{
 \includegraphics[width=2.6in,height=2.0in, angle=0]{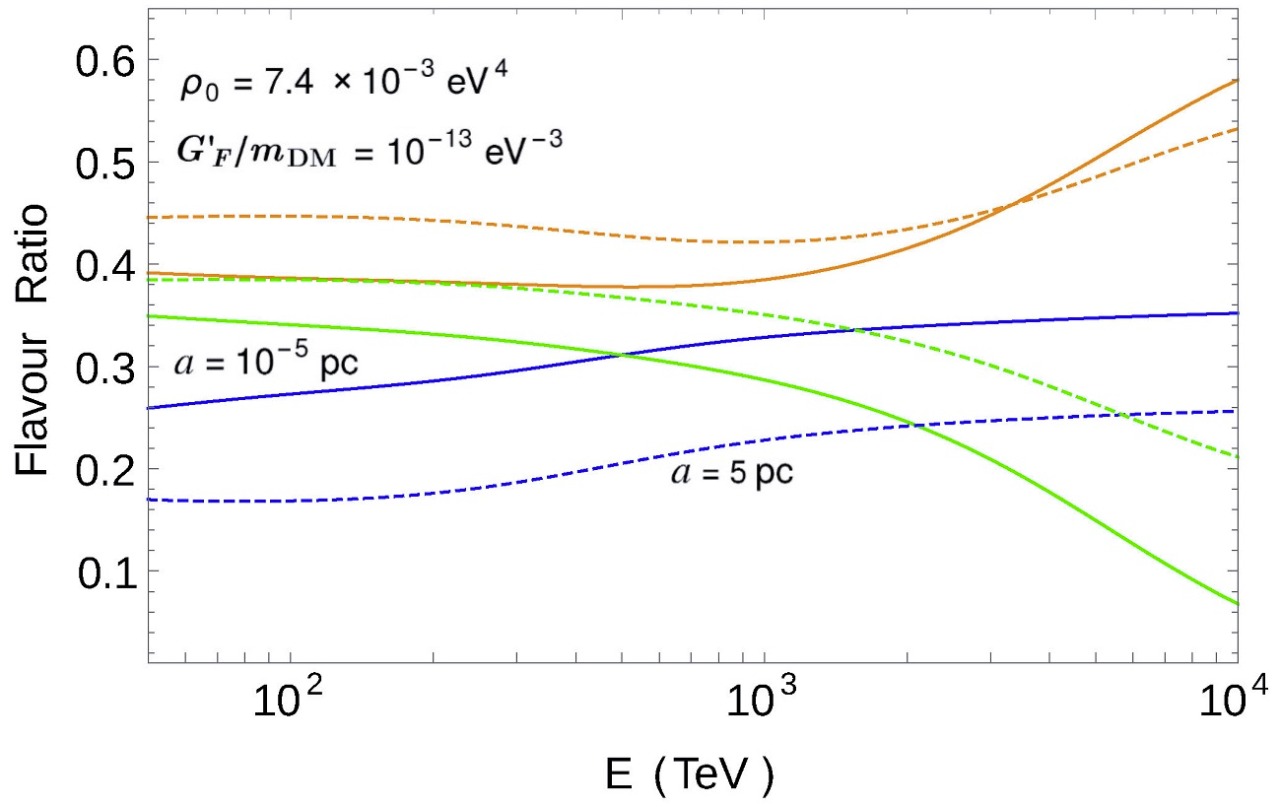}}   
 \caption{Energy dependence of flavour ratios  at $z=2$ for  (a) neutrinos ($f^{D}_{\a}$) and (b) antineutrinos ($f^{D}_{\bar{\a}}$), and (c) the average of neutrinos and antineutrinos ($F^{D}_{\a}$). The blue, orange and green lines represent flavour ratios for $e$, $\mu$ and $\tau$ flavours. Solid and dashed  lines stand for $a=10^{-5}$~pc~(non-adiabatic case) and $a = 5$~pc~(adiabatic case) respectively.}
 \label{fig2}
\end{center}
 \end{figure} 

\begin{itemize}
\item For $E<1$~PeV,  both neutrinos and antineutrinos undergo adiabatic transition for $a \gtrsim 10^{-3}$~pc. Hence, the flavour ratios at the detector can be written as
\bea
f_{\beta}^D &=& P_{\alpha \beta} f_{\alpha}^S=|U_{\beta i}^D|^2 |U_{\alpha i}^S|^2 f_{\alpha}^S = |U_{\beta i}^D|^2 |U_{e i}^S|^2+ 2|U_{\beta i}^D|^2 |U_{\mu i}^S|^2.
\eea 
As $U^S$ tends to an identity matrix at high energies, the above relation simplifies to $f_{\beta}^D= |U_{\beta 1}^D|^2+ 2|U_{\beta 2}^D|^2$. With increasing energy, the off-diagonal terms of $U^D$ decrease as well. Thus the fraction of $\nu_{\mu}$ increases and $\nu_{\tau}$ nearly diminishes at higher energies. As a result,  the flavour ratio at the detector becomes $1:2:0$ at energies $E \gtrsim 50$~PeV, the same as the flavour ratio at source.
 This can be seen in fig.~\ref{fig2} for $a = 5$~pc, which corresponds to the adiabatic case.  As it can be seen from eq.~\eqref{a}, such a value of $a$ can be  achieved for $\mbh \sim 10^6~\rm{\mso}$ and $\mdm \sim 1.3 \times 10^{-20}$~eV.
  Further, as will be discussed in details at the end of this section, $\rho_0 > 10^{-3}$~eV$^4$ leads to $U_S \sim I$. Hence, any value of $\rho_0$ greater than $10^{-3}$~eV$^4$ leads to the same flavour ratio as shown in fig.~\ref{fig2}, keeping the other parameters same. 
The adiabatic propagation can also be achieved in the scenarios where the BH does not dominate the source~\cite{Davies:2019wgi}. 
 
For $\mbh < \mh$ the dark matter density profile at the core is given by~\cite{Davies:2019wgi,Schive:2014hza}
\begin{align*}
\rho(r) &= \frac{0.019 (\mdm/10^{-22}\text{eV}) (r_c/\text{kpc})^{-4}}{[1+0.091 (r/r_c)^2]^8} \mso \text{pc}^{-3}, \\
r_c &= \frac{2.2 \times 10^{8} (\mdm/10^{-22}\text{eV})^{-2}}{(M_{\text{sol}}/\mso)}~\text{kpc}
\end{align*}
where $r_c$ is the radius of the core which contains the maximum mass of the soliton.
 Thus, the radial change in DM number density is $|\text{d} \ln n_{\phi}/\text{d} r| = (1.45 r/r_{c}^2)/[1+0.091 (r/r_c)^2]$.
As mentioned before, for $\mbh < \msol$ the core-halo relation is presented in ref.~\cite{Schive:2014hza}. 
  For $\mdm=10^{-22}$~eV and $\mh = 10^9 \mso$, one obtains $r_c \sim 1$~kpc, $\rho(0) \sim 5 \times 10^{-7}$~eV$^4$, and $|\text{d} \ln n_{\phi}/\text{d} r| \lsim 1.33$~kpc$^{-1}$ for $r \gtrsim 1$~kpc. For such a value of $|\text{d} \ln n_{\phi}/\text{d} r|$, adiabatic propagation is guaranteed below $\mathcal{O}(10^6)$~PeV.

\item In the case of neutrinos, for the benchmark $a = 10^{-5}$~pc, there is a significant violation of adiabaticity only for the jumping of $\nu_{1}$ to $\nu_{3}$ for $E \gtrsim 1$~PeV. Hence, the ratio of electron neutrinos at the detector is given as
\begin{align*} 
f_{e}^D= |U_{ei}^D|^2 |U_{\alpha i}^S|^2 f_{\alpha}^S-P^{c}_{31}(|U_{e1}^D|^2-|U_{e3}^D|^2)(|U_{\alpha 1}^S|^2-|U_{\alpha 3}^S|^2) f_{\alpha}^S.
\end{align*}
As mentioned earlier, for fixed value of $\rho_0$  and $G_F'$, as the energy increases the effective Hamiltonian is dominated by the matter potential term. Thus the mixing matrices $U_{S}$ and $U_{D}$ tend to identity matrix leading to $f_{e}^D= (1-P^{c}_{31}) f_{e}^S$. Taking $f_{\alpha}^S=(1:2:0)$,  $f_{e}^D$ equals $ (1-P^{c}_{31})$.
For the benchmark shown in fig.~\ref{fig2}(a) with $a = 10^{-5}$~pc, $P^{c}_{31}$ increases with energy and finally attains a constant value of $P^{c}_{31} \sim 0.98$. As a result, $f_{e}^D$ decreases with energy and eventually saturates. For extremely non-adiabatic propagation the limiting value of the crossing probability is $P_{ij}^{c} = \cos^2 \theta_{ij}$.
Similarly, for $f_{\alpha}^S=(1:2:0)$ the ratio of muon neutrinos at high energies simplifies to, 
\bea
f_{\mu}^D &=& 2|U_{\mu 2}^D|^2 |U_{\mu2}^S|^2 -2P^{c}_{31}(|U_{\mu1}^D|^2-|U_{\mu 3}^D|^2)(|U_{\mu1}^S|^2-|U_{\mu 3}^S|^2)\nn\\
&& - P^{c}_{31}(|U_{\mu1}^D|^2-|U_{\mu 3}^D|^2)(|U_{e1}^S|^2-|U_{e3}^S|^2) \nn \\
&\simeq & 2-P^{c}_{31}(|U_{\mu1}^D|^2-|U_{\mu 3}^D|^2). \nn
\eea
For this benchmark, the combination of the off-diagonal elements $(|U_{\mu1}^D|^2-|U_{\mu 3}^D|^2)$ decreases with increasing energy. Thus, $f_{\mu}^D$ increases with energy. Also, the fraction of tau neutrinos at high energies simplifies to, 
\bea
f_{\tau}^D &=& |U_{\tau 1}^D|^2 |U_{e1}^S|^2+2|U_{\tau 2}^D|^2 |U_{\mu2}^S|^2 -2P^{c}_{31}(|U_{\tau 1}^D|^2-|U_{\tau 3}^D|^2)(|U_{\mu1}^S|^2-|U_{\mu 3}^S|^2)\nn\\
&&-P^{c}_{31}(|U_{\tau 1}^D|^2-|U_{\tau 3}^D|^2)(|U_{e1}^S|^2-|U_{e3}^S|^2) \nn \\
&\simeq & - P^{c}_{31}(|U_{\tau 1}^D|^2-|U_{\tau 3}^D|^2). \nn
\eea
Subsequently, as it can be seen from fig.~\ref{fig2}(a), for $a = 10^{-5}$~pc,  the flavour ratio for neutrinos tend towards $0:2:1$ at higher energies.
 Using eq.~\eqref{a}, for $\mbh \sim 8 \times 10^{6}~\mso$ and $\mdm \sim 3.16 \times 10^{-18}$~eV one obtains  $a \sim 1.1 \times 10^{-5}$~pc. This is a typical scenario which leads to non-adiabatic flavour transition for both neutrinos and antineutrinos.  In this paper we consider the source to be located at $z = 2$.

\item The flavour ratios for electron and muon antineutrinos after non-adiabatic transition can be simplified to  
\begin{align*} 
\frac{f_{\bar{e}}^D}{f_{\bar{\mu}}^D}=\frac{|U_{e1}^D|^2+ 2 |U_{e2}^D|^2-2 P^{c}_{32}(|U_{e2}^D|^2-|U_{e3}^D|^2)}{2|U_{\mu 2}^D|^2(1-   P^{c}_{32})}.
\end{align*}
With increasing energy the off-diagonal terms in $U^D$ decrease and $P^{c}_{32}$ increases before it finally saturates. As an example, for $a = 10^{-5}$~pc,  $P^{c}_{32}$ attains a value of $\sim 0.43$ for $E \gtrsim 3$~PeV. Thus, $f_{\bar{e}}^D$ increases and $f_{\bar{\mu}}^D$ decreases at higher energies, leading to $f_{\bar{e}}^D/f_{\bar{\mu}}^D=1/(2-2 P^{c}_{32}) \sim 1$ for $E \gtrsim 10$~PeV. This has been  shown in fig.~\ref{fig2}(b).    

\end{itemize}

 At the detector, DM effective potential is evaluated as $V^{D}_{\tau\tau} = \gfp \rho_D/\mdm$ with $\rho_D \simeq 0.4$~GeV cm$^{-3}$. Keeping $\gfp/\mdm= 10^{-13}$~eV$^{-3}$, the resonance condition in  eq.~\eqref{eq:resonance} implies that the resonance energy at the detector is $E^R_{31}=3.8$~PeV. Following eq.~\eqref{eq:mattermixing} the mixing angles at the detector tend to those in the vacuum for $E < 3.8$~PeV.

\begin{figure}[h!]
 \begin{center}
\subfigure[]{
 \includegraphics[width=2.6in,height=2.0in, angle=0]{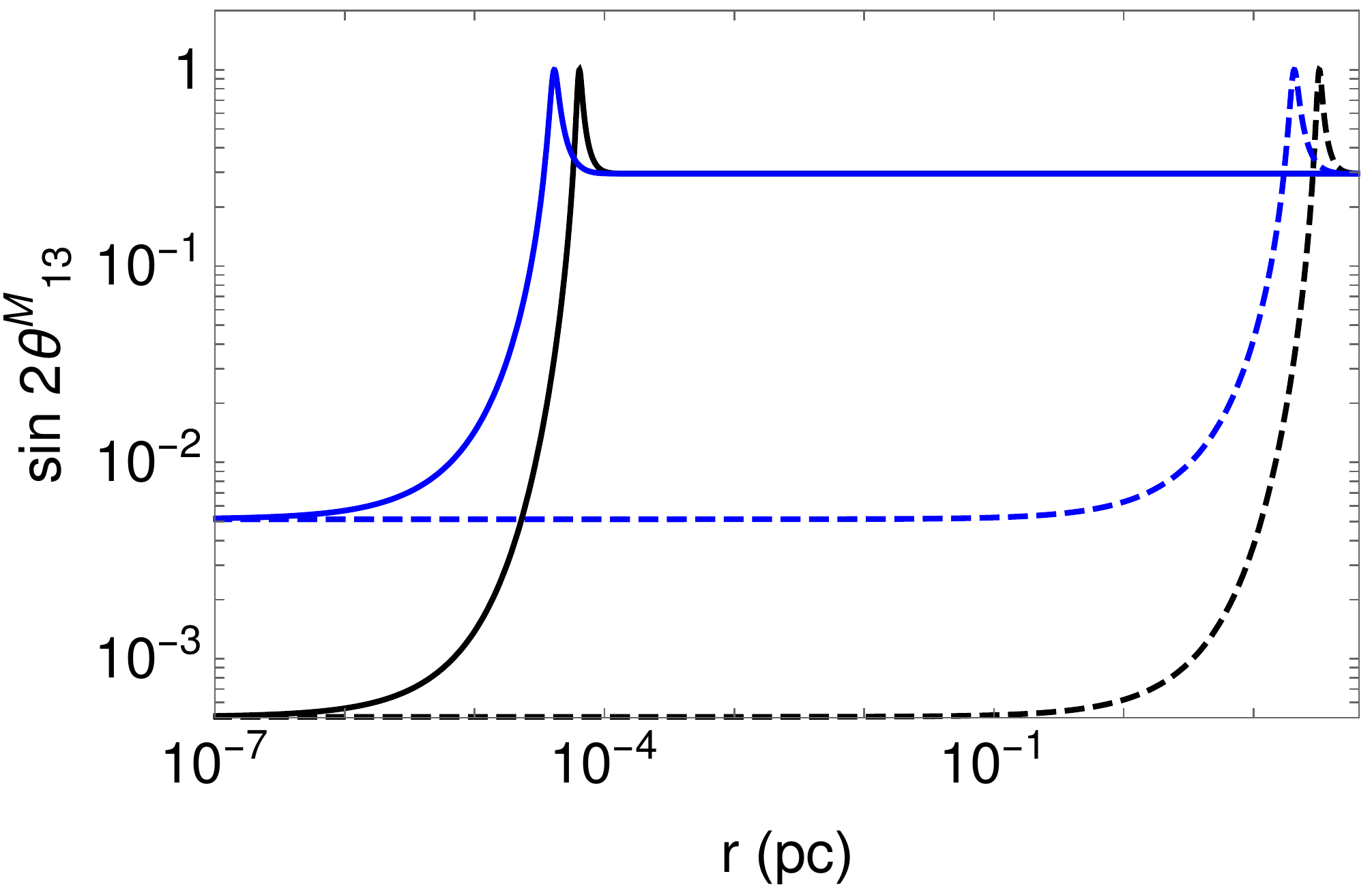}}
 \hskip 15pt
 \subfigure[]{
 \includegraphics[width=2.6in,height=2.0in, angle=0]{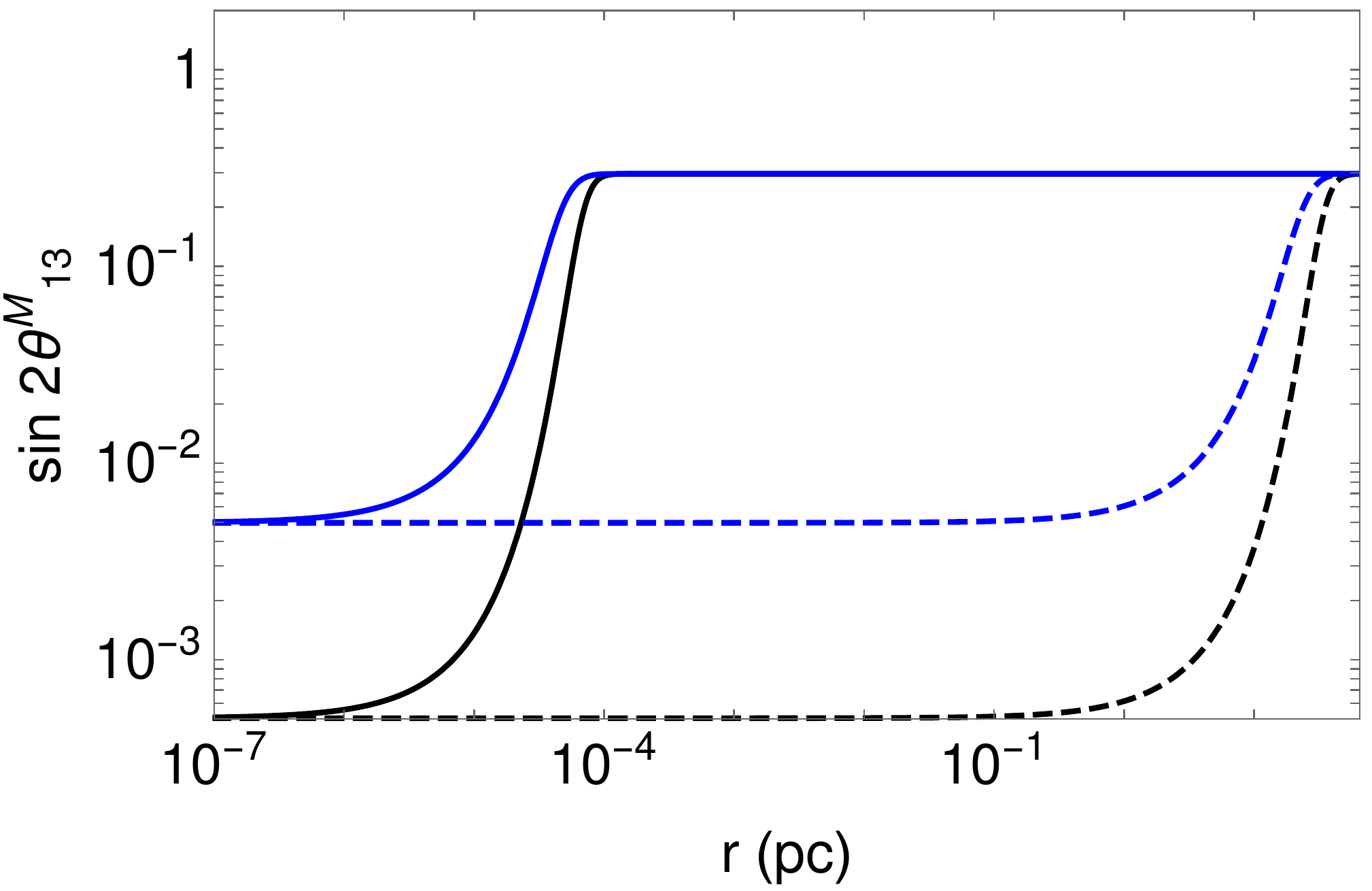}}
 \subfigure[]{
 \includegraphics[width=2.6in,height=2.0in, angle=0]{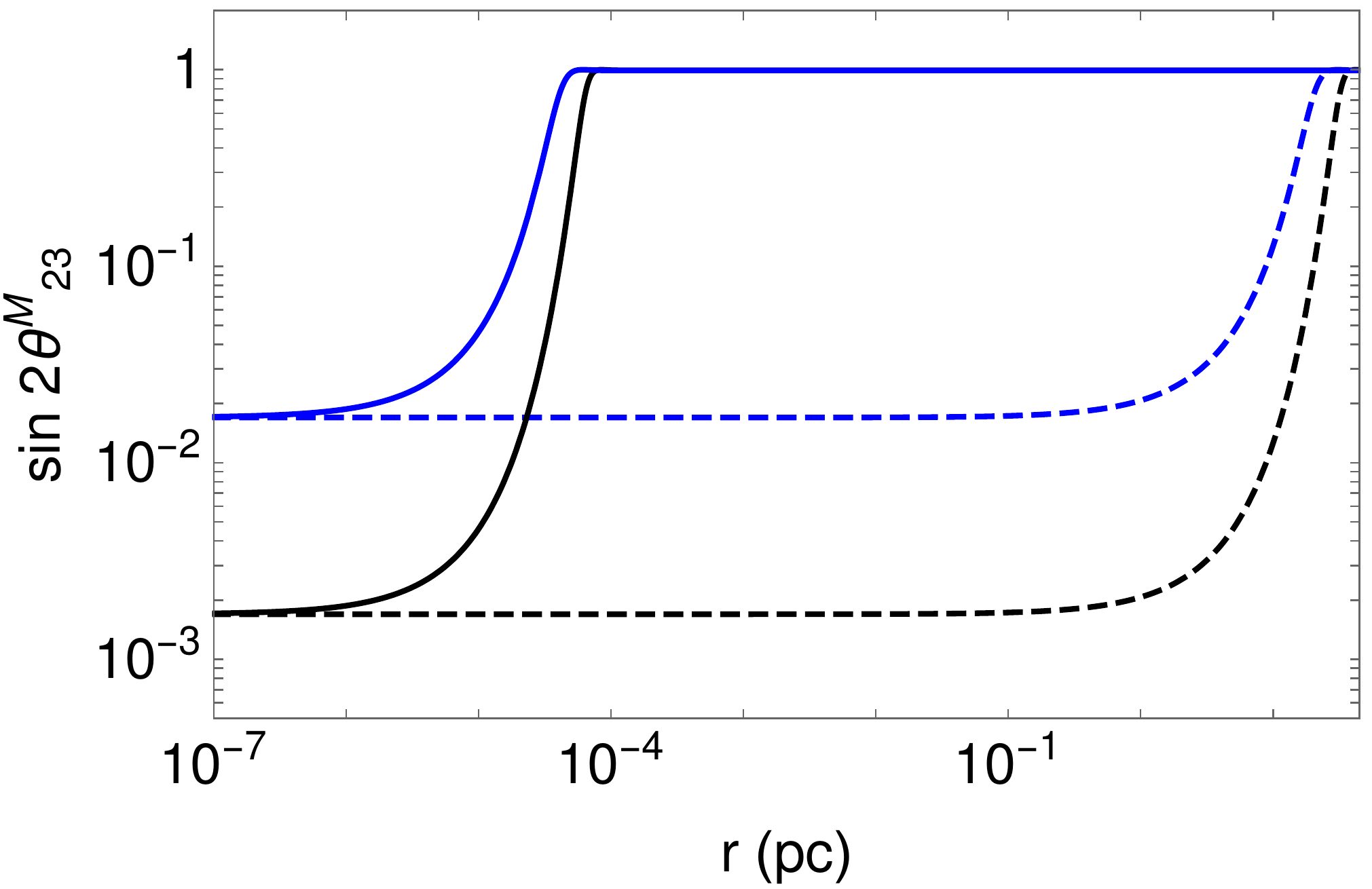}}   
 \caption{$\sin 2 \theta_{M}$ vs. radial distance at source with $\rho_0=7.4 \times 10^{-3}$~eV$^{4}$ and $\gfp/\mdm= 10^{-13}$~eV$^{-3}$ for (a) positive $V_{\tau \tau}$, (b) negative $V_{\tau \tau}$ and (c) negative $V_{\tau \tau}$. The dashed~(solid) line represents $a=5$ pc~($10^{-5}$~pc), whereas the black~(blue) line represents  $E_S=1$~PeV (100~TeV). Here, $E_S$ is the energy at the source situated at $z=2$.} 
 \label{fig3a}
\end{center}
 \end{figure}

  At source, the steep fall in the DM density assures that, for the considered energy range, there always exists a point at the source where the density ($\rho_R$) meets the resonance condition for $\theta_{13}$,  given $V_{\tau \tau} >0$ and $\rho_0 > \rho_R$, as shown in fig.~\ref{fig3a}~(a). In fig.~\ref{fig3a}~(a) the position where resonance occurs can be read from the peak in $\sin 2 \theta^M_{13}$. Here, $E_S= E (1+z)$ is the neutrino energy at source for $z=2$.  Similarly, for the energy under consideration, $E \in [0.05, 10]$~PeV, $\theta_{23}$ undergoes resonance at source, given $V_{\tau \tau} <0$ and $\rho_0 > \rho_R$, as shown in fig.~\ref{fig3a}~(c). Fig.~\ref{fig3a}~(b) depicts the case where $V_{\tau \tau} <0$, hence, the resonance condition is never met for $\theta^M_{13}$ and $\theta^M_{13} \sim \theta_{13}$ due to lower DM density at source.

\begin{figure}[h!]
 \begin{center}
 \includegraphics[width=2.6in,height=2.0in, angle=0]{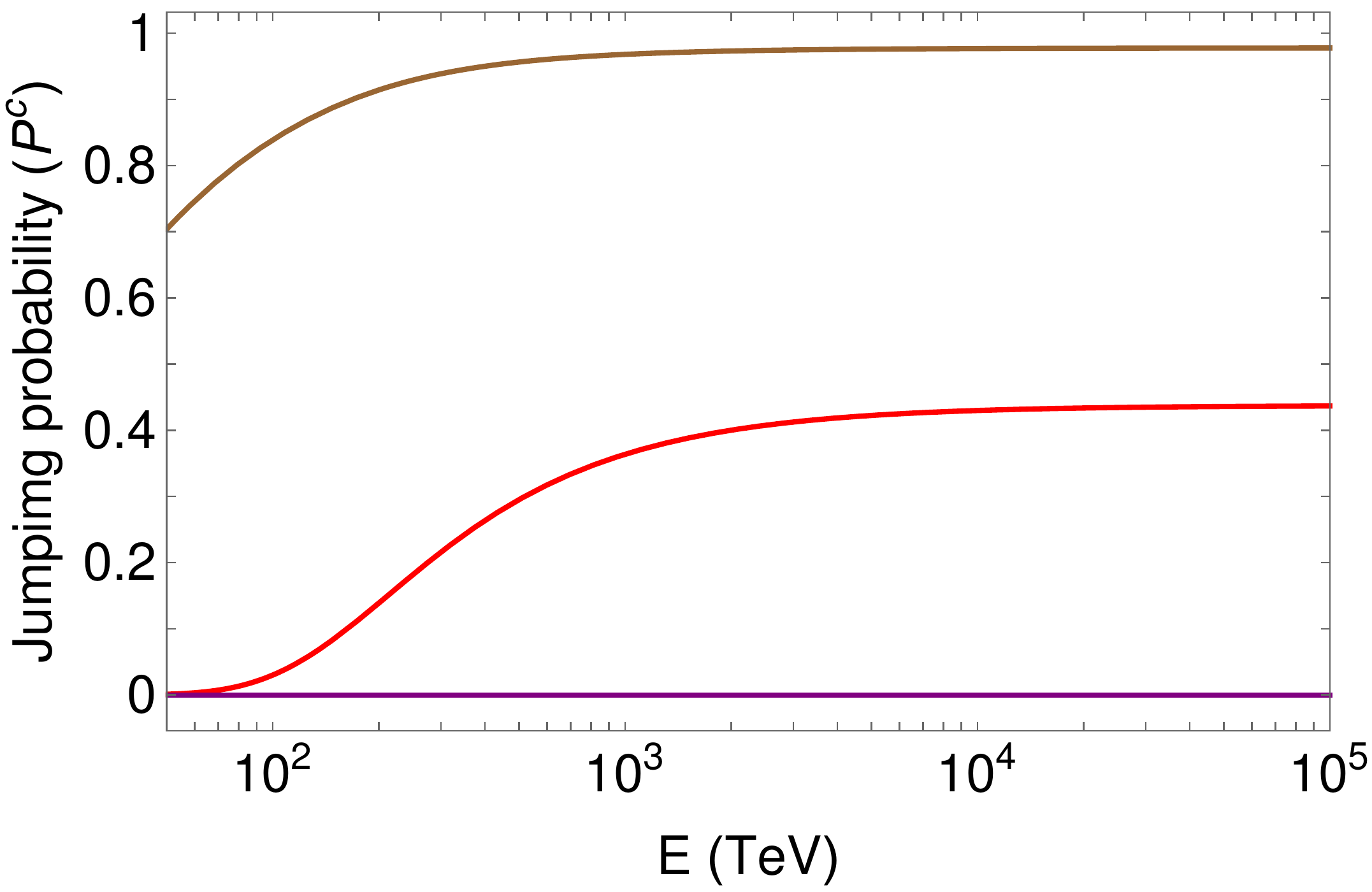}
\caption{Jumping probability $P^c_{ij}$ from mass state $i$ to $j$ at resonance vs. neutrino energy.  The brown and red lines represent $P^{c}_{31}$ and  $P^{c}_{32}$ for $a=10^{-5}$~pc respectively. The purple line represents  $P^{c}_{3i}$  $(i=1,2)$ for $a=5$~pc, $z=2$.}
 \label{fig4a}
\end{center}
 \end{figure}

 Also, the jumping probability between the mass states increases with the decrease in adiabaticity parameter, which in turn, is proportional to $\Delta m^2_{ij} \sin^2 2 \theta_{ij}$ and $1/( E | \text{d} \log \rho / \text{d} r|_R)$. Hence, $a < 10^{-2}$~pc leads to large $| \text{d} \ln \rho / \text{d} r|_R$. The jumping probability is significant for such values of $a$, as shown in fig.~\ref{fig4a}. 
 Therefore, in the extreme non-adiabatic regime, the crossing probability is significant at the position where resonance occurs at the source. At high energy, the energy dependence of the flavour ratio stems from the crossing probability. At lower energies, propagation is supposed to be adiabatic, but the mixing angles in the densities $\rho_0$ and $\rho_D$ are different, leading to changes in the flavour compositions of the observed neutrinos.

\begin{figure}[h!]
 \begin{center}
\subfigure[]{
 \includegraphics[width=3in,height=2.3in, angle=0]{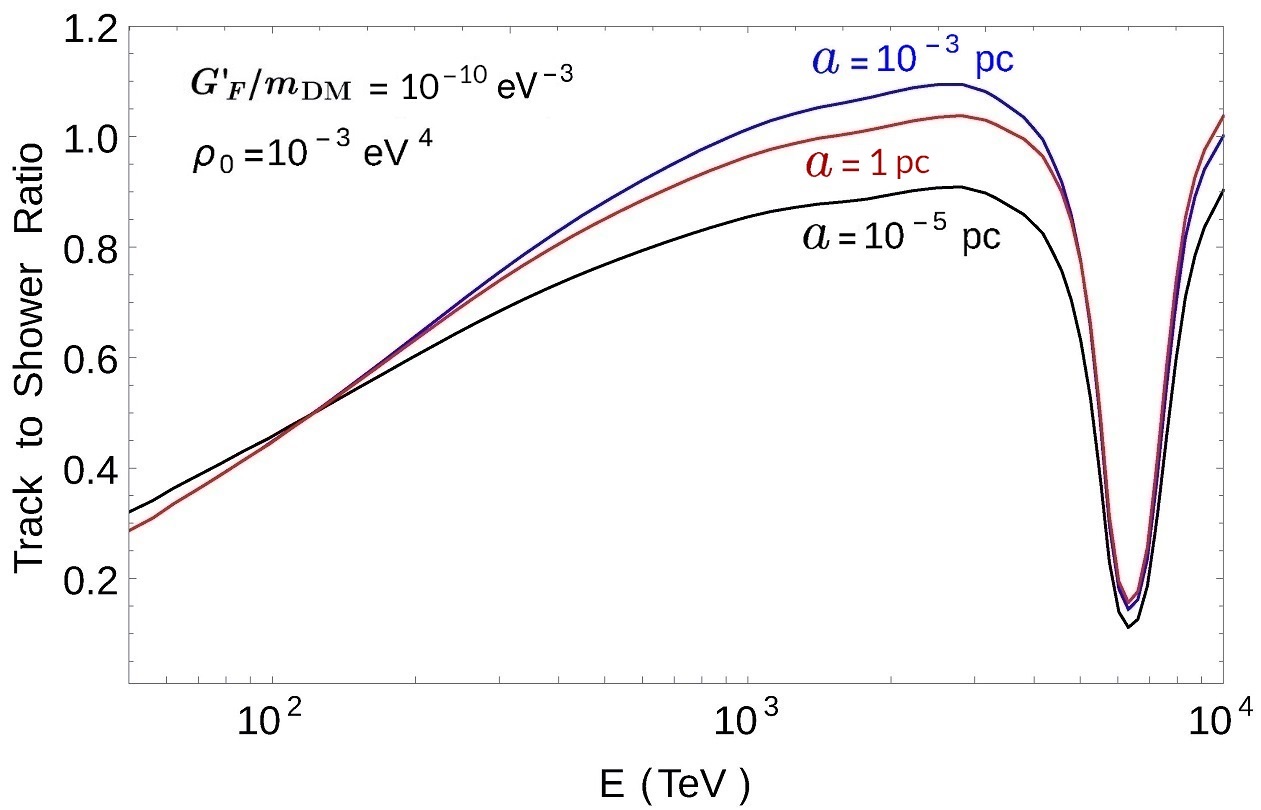}}
 \hskip 15pt
 \subfigure[]{
 \includegraphics[width=3in,height=2.3in, angle=0]{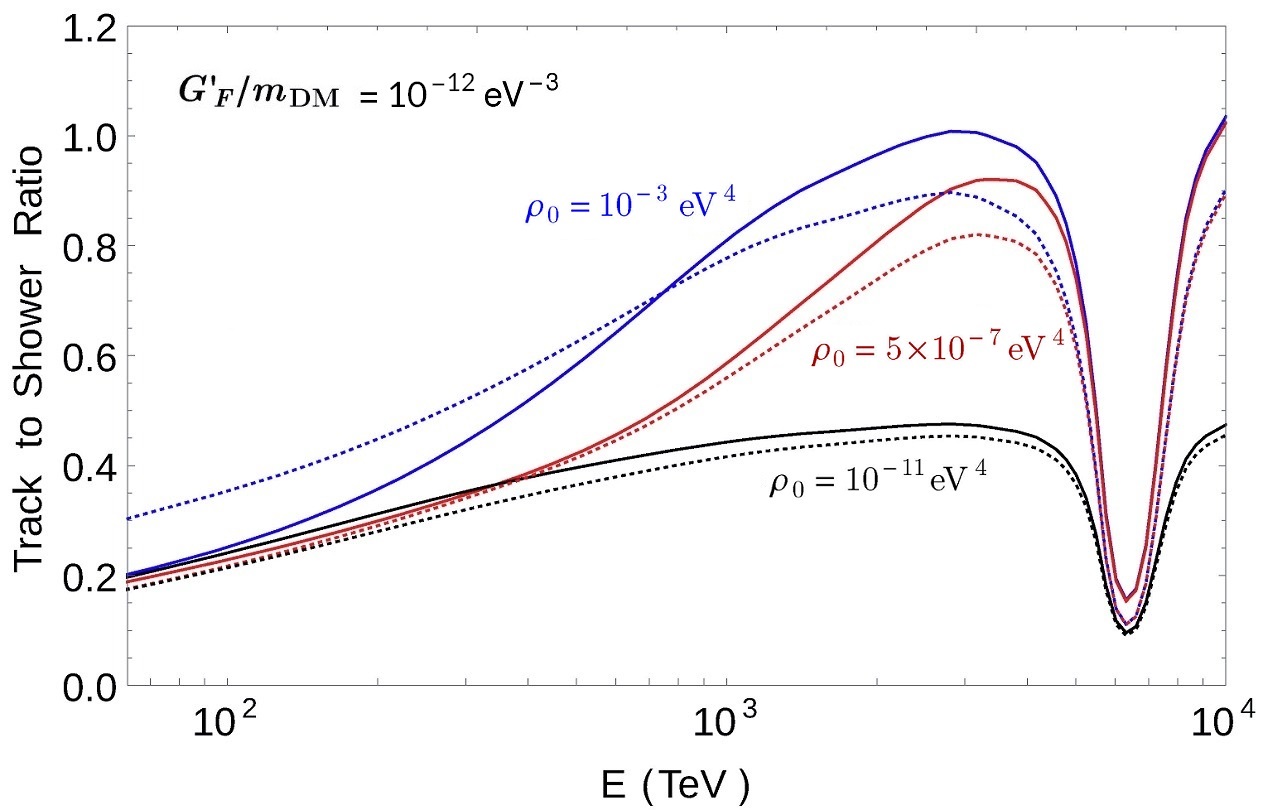}}
 \hskip 15pt
 \subfigure[]{
 \includegraphics[width=3in,height=2.3in, angle=0]{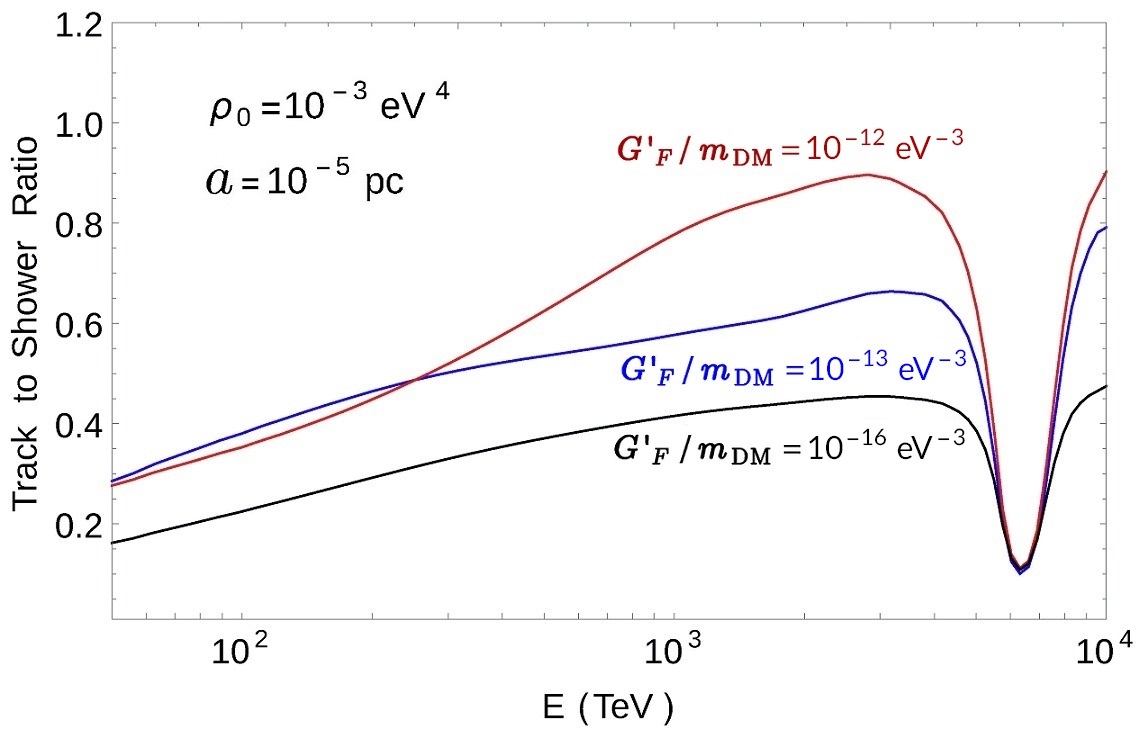}}
 \caption{Track to shower ratio vs. neutrino energy at $z=2$ for  varying values of (a) radius of the solitonic core $a$, (b) DM density at its centre $\rho_0$, and (c) $G'_F/\mdm$. In (b), the solid and dotted lines correspond to $a = 1$~pc~(adiabatic case) and $10^{-5}$~pc~(non-adiabatic case) respectively.   }
 \label{fig4}
\end{center}
 \end{figure}

At IceCube, the flavour ratios of astrophysical neutrinos are extracted from the track to shower ratio
\bea
\frac{N_{\text{track}}}{N_{\text{shower}}}= \frac{0.8 A_{\mu} f_{\mu}^D+ 0.13 A_{\tau} f_{\tau}^D }{A_{e} f_{e}^D+0.2 A_{\mu} f_{\mu}^D+ 0.87 A_{\tau} f_{\tau}^D},
\label{tr2sh}
\eea
where $A_{l}$ is the effective area for detecting neutrinos of flavour $l$ provided by IceCube collaboration~\cite{Aartsen:2013jdh}. In eq.~\eqref{tr2sh} we use that, the probabilities of obtaining a track from a $\nu_{\mu}$ and $\nu_{\tau}$ are approximated as 0.8 and 0.13 respectively~\cite{Palladino:2015zua}. 

At energies lower than $\sim$ 100~TeV both electromagnetic shower produced by $\nu_e$ and hadronic shower by $\nu_{\tau}$ lead to cascade signatures at IceCube~\cite{Abbasi:2020zmr}.
Thus, at lower energies, track to shower ratio is the only probe of the flavour ratios. 
But for energies higher than $\sim$~PeV, $\nu_{\tau}$  can leave distinguishable signatures in the form of  double bang and lollipop events,  with visible tau tracks~\cite{Cowen:2007ny}. 
Moreover, it has been suggested that, hadronic showers can be distinguished from electromagnetic showers in the TeV-PeV range by means of `pion and neutron echos'~\cite{Li:2016kra}.
If such a distinction of electron and tau neutrino events are successfully performed, the flavour ratio can be known with an unprecedented accuracy after a substantial livetime of IceCube-Gen2 operation~\cite{MB}.

In the following, we discuss the change in the track to shower ratio for varying values of $a$, $\rho_0$, and $G'_F$ shown in fig.~\ref{fig4}. Note that, we also take into account the contributions of antineutrinos while computing the track to shower ratio. The effective areas for neutrinos and antineutrinos are the same except for the electron flavour due to the possibility of hadronic shower induced by $\bar{\nu}_{e}$ at $E \sim 6.3$~PeV.   

\pagebreak

\begin{itemize}

\item {\bf Dependence on $a$}\,:

In fig.~\ref{fig4}(a) we have shown the changes in track to shower ratio for $a = 10^{-5}$~pc, $10^{-3}$~pc, and 1~pc with fixed values of $\rho_{0}= 10^{-3}$~eV$^4$ and $G_{F}^{\prime}/\mdm = 10^{-10}$~eV$^{-3}$. 
As mentioned in Sec.~\ref{sec:III}, with the neutrino energies considered in this paper,  both neutrinos and antineutrinos undergo adiabatic transition for $a=1$~pc. Similarly, for $a = 10^{-5}$~pc, both neutrinos and antineutrinos undergo non-adiabatic transitions. 
However, for $a = 10^{-3}$~pc, only neutrinos can have non-adiabatic transition.

As can be seen from fig.~\ref{fig2}(b), for  $a=10^{-5}$~pc, $f_{\bar{\mu}} \sim f_{\bar{e}}$ and $f_{\bar{\tau}} \sim 0$ at higher energies. On the other hand, for $a=10^{-3}$~pc the antineutrinos propagate adiabatically leading to the flavour ratio $\sim 1:2:0$ at high energies. Thus, the value of track to shower ratio is larger for $a=10^{-3}$~pc compared to the case of $a=10^{-5}$~pc. 

Following the previous logic, one may apparently expect the track to shower ratio to be higher for $a=1$~pc compared to $a = 10^{-3}$~pc. 
But, as it was shown in fig.~\ref{fig2}(a),  the ratio $f_{\mu}^D/(f_{e}^D + f^D_{\tau})$ is higher for the non-adiabatic case compared to the adiabatic case. For $a = 10^{-3}$~pc, neutrinos oscillate non-adiabatically, thereby leading to a higher value of track to shower ratio compared to $a = 1$~pc.  

\item {\bf Dependence on $\rho_0$}\,:

To understand the dependence of track to shower ratio  on DM density at the source, in fig.~\ref{fig4}(b) we consider three benchmark values $\rho_{0}= 10^{-3}$~eV$^4$,  $5\times10^{-7}$~eV$^4$, and $10^{-11}$~eV$^4$, while fixing $a= 1$~pc~($10^{-5}$~pc), $G_{F}^{\prime}/\mdm=10^{-12}$~eV$^{-3}$ and $z=2$. 
These three benchmark values of $\rho_0$  lead to resonance energies at source $E_{31}^R \big|_{source} \sim 1.2 \times 10^{12}$~eV, $2.4 \times 10^{15}$~eV, $1.2 \times 10^{20}$~eV, respectively.

Let us first discuss the adiabatic case with $a = 1$~pc.
 For $E \lsim E_{31}^R$, the vacuum term in the effective Hamiltonian is more significant than the matter term.
 Thus, in the limit $E \ll E^{R}_{ij}$, the flavour ratio is close to $1:1:1$.
 Though, at much higher energies the matter term becomes more significant, resulting in a flavour ratio of $\sim 1:2:0$, as it is expected in a typical adiabatic scenario. 
 This was also shown in fig.~\ref{fig2}(c) for $a = 5$~pc, where the ratio of averaged muon-flavour contribution~($F^D_{\m}$) to the electron and tau (anti)neutrinos increase with energy. 
  Thus, the case with a lower value of $E^R$ will lead to a larger fraction of muon (anti)neutrinos. 
  So, as can be seen from fig.~\ref{fig4}(b), the benchmark with higher DM density at source has a larger value of track to shower ratio till the resonance energy.
 
The non-adiabatic case also shows similar features. Note that, the track to shower ratio becomes saturated to its maximum value for $\rho_0 \gtrsim 10^{-3}$~eV$^4$ and to its minimum value at $\rho_0 \lesssim 10^{-11}$~eV$^4$. In other words, the case with $\rho_0 = 10^{-11}$~eV$^4$  almost coincides with the standard scenario with no DM at the source.

\item {\bf Dependence on $G'_F/\mdm$\,}:

 In fig.~\ref{fig4}(c) we consider $G_{F}^{\prime}/\mdm = 10^{-16}$~eV$^{-3}$, $10^{-13}$~eV$^{-3}$, and $10^{-12}$~eV$^{-3}$, with a fixed DM  profile $\rho_{0}= 10^{-3}$~eV$^4$ and $a=10^{-5}$~pc. 
As it can be seen from fig.~\ref{fig2}(c) at lower energies,  though the value of $F^{D}_{\mu}$ is higher than the individual electron or tau-flavour contributions, the value of $F^{D}_{\mu}/(F^{D}_{e} + F^{D}_{\tau})$ is slightly lower than one. But for $E \gtrsim 500$~TeV,  the muon-flavour contribution $F^{D}_{\mu}$ increases with energy and the tau-flavour contribution abruptly decreases, leading to $F^{D}_{\mu}/(F^{D}_{e} + F^{D}_{\tau}) > 1$. 

The resonance energy is inversely proportional to $V_{\tau \tau}$, and therefore to, $G_{F}'/\mdm$. Thus, the flavour ratios for $G_{F}^{\prime}/\mdm = 10^{-12}$~eV$^{-3}$, $f_{l}(E)$ can be related to the flavour ratios for $G_{F}^{\prime}/\mdm = 10^{-13}$~eV$^{-3}$, $f'_{l}(E)$, such that $f_l(E) \simeq f'_l(E/10)$.
 Hence, as can be inferred from fig.~\ref{fig2}(c), for $E \gtrsim 300$~TeV the case with $G_{F}^{\prime}/\mdm = 10^{-12}$~eV$^{-3}$ has a higher muon (anti)neutrino contribution, and in turn, a larger value of track to shower ratio. Also it can be seen from fig.~\ref{fig2}(c) that, for lower energies, $F^{D}_{e}$ slightly increases whereas $F^{D}_{\mu}$ and $F^{D}_{\tau}$ do not significantly change. Thus, $G_{F}^{\prime}/\mdm = 10^{-12}$~eV$^{-3}$ leads to a larger value of $F^{D}_{e}$, and therefore a smaller value of track to shower ratio. These effects can be read off fig.~\ref{fig4}(b). 
Moreover, for $G'_F/\mdm = 10^{-16}$~eV$^{-3}$, the track to shower ratio attains its standard value. 

\end{itemize}

In this paper, we have considered only positive values of $V_{\tau \tau}$. Although, for negative values of $V_{\tau \tau}$, the ratio $f^{D}_{\bar{\mu}}/(f^D_{\bar{e}}+f^D_{\bar{\tau}})$ becomes larger compared to the case of positive $V_{\tau\tau}$.
 Hence, one can expect a larger value of the track to shower ratio near $E  \sim 6.3$~PeV. 
 
As it was pointed out in Sec.~\ref{sec:III}, $\epsilon_{\mu\mu}$ is  constrained at $\mathcal{O}(10^{-26})$~eV$^{-2}$. Thus, non-zero values of $V_{\mu\mu}$ may also lead to significant changes in the track to shower ratios for rather small values of $\mdm$. Also, scalar-mediated neutrino-DM effective interactions are constrained from invisible $Z$ decay at $\mathcal{O}(10^{-10})$~eV$^{-1}$~\cite{Pandey:2018wvh} and may lead to observable effects in flavour ratios. Though, in a concrete model with a $Y=2$ triplet scalar, the smallness of neutrino mass renders this effective interaction to be quite small to have any interesting effect on flavour ratios.

The relations between the SMBH, halo, and soliton masses may not be valid for the range of $\mdm \gtrsim 10^{-22}$~eV~\cite{Davies:2019wgi, Chavanis:2019amr, Bar:2019pnz}. As a consequence, in this paper, we  consider $\rho_0$ and $a$ as the parameters describing DM profile while demonstrating the energy dependence of the flavour ratios. These issues are under active scrutiny and a better understanding of the interplay between $\mbh$, $\msol$, and $\mh$ is expected to emerge in the future, which will further strengthen our prescription by reducing the number of unknown variables in the fit.

\section{Summary and outlook}
In the astrophysical neutrino sources, such as AGNs, the matter accretion disc and a dark matter halo can surround a super-massive black hole.
Sub-eV ultralight scalar dark matter, in a form of Bose-Einstein condensate, happens to be a suitable candidate for cold dark matter. An interaction of these high energy neutrinos with  such ultralight DM is an interesting proposal that helps to address various features of the observed neutrino spectrum, as well as the lack of directional coherence with particular astrophysical objects. Like AGNs, GRBs can be located at regions with a significant DM density~\cite{Huang:2020twt} and can  lead to changes in  flavour ratios at IceCube. In that spirit, in this paper we have considered if such interactions can be important for neutrino astronomy through the observation of neutrino flavour ratios at earth.  

We find that while passing through the DM halo, the details of the halo profile, DM mass, the redshift associated with the AGN, the strength of such interactions, masses of the SMBH and DM halo get encoded into the energy dependence of neutrino flavour ratio at the IceCube. In future, the statistics at IceCube will improve with the Gen2 upgrade. 
Beside IceCube,  KM3NeT/ARCA will also have the potential to detect point-like extragalactic neutrino sources~\cite{Adrian-Martinez:2016fdl}. 
The accuracy for directionality in ARCA can even be better compared to the IceCube, making it a somewhat better probe of  such astrophysical sources~\cite{Aiello:2018usb}. 
 At that point, more such neutrinos can possibly be traced back to the potential astrophysical sources. 
This will allow the usage of our proposed method to perform neutrino astronomy. The knowledge of some of the parameters like the masses of the SMBH and DM halo, the distance of the AGN from other modes of astronomy may help us improve the fit to the rest of the unknowns.

In spite of the large neutrino energies, the centre of mass energy for neutrino scattering off ultralight DM  is much less compared to the mass of the particle mediating such interactions. The latter can be of  $\mathcal{O}$(MeV) for the case of a light $Z^\prime$.
In such cases, the resulting $\nu$-DM cross-section is negligible, to lead to any appreciable neutrino flux suppression~\cite{Pandey:2018wvh}. We have shown that, even for $\nu$-DM interactions feeble enough to impart any changes in astrophysical neutrino flux, the track to shower ratios can significantly modify due to the large DM number density.   
 Even if the nature of effective interactions are more complicated with additional momentum dependencies, the mass terms for neutrinos vary only by factors of $\mdm$. This allows us to define a single effective interaction strength $G_F^\prime$. 

In this paper, we have used the current best-fit values of neutrino mass and mixing parameters for normal ordering. In the case of inverted ordering, both $\Delta m_{31}^2$ and $\Delta m_{32}^2$  become negative, whereas the mixing angles remain in the same octants. Thus, for $ij = 32$ and $ij = 31$ the resonance condition $2 E^R_{ij}\, V_{\tau\tau} = \Delta m_{ij}^2 \cos 2 \theta_{ij}$ is satisfied for a positive and negative value of $V_{\tau \tau}$ respectively. Note that, so far we have restricted our discussion to the best-fit values of $\Delta m_{ij}^2$ and $\theta_{ij}$. But, for both the normal and inverted ordering, $3 \sigma$ allowed ranges on $\theta_{23}$  span over the first and second octants, thereby allowing both positive and negative values of $\cos 2\theta_{23}$. So, deviating from the best-fit values opens up the possibility of resonant effects in both 32 and 31 transitions. As mentioned in Sec.~\ref{sec:III}, this effect leads to new terms proportional to $P^c_{32} P^c_{31}$ in the final flavour ratios. However, future reactor experiments, such as JUNO~\cite{An:2015jdp}, atmospheric neutrino experiments like HyperK~\cite{Abe:2018uyc}, PINGU~\cite{Aartsen:2014oha}, and accelerator experiments like DUNE~\cite{Acciarri:2015uup}, T2HK/T2HKK~\cite{Abe:2016ero}  aim to resolve neutrino mass ordering and octant degeneracy. Certain combinations of these experimental data, for example, JUNO+PINGU~\cite{Bezerra:2019dao}, Daya Bay II+PINGU~\cite{Blennow:2013vta} can be decisive for this purpose through synergy effects. With a better understanding of the mass ordering and the sign of $\cos 2\theta_{23}$, the determination of $G'_F$ within our framework can be easier.

Another important aspect is the role of local DM density in this proposal. The galactic DM density does not change that rapidly to lead to any non-adiabaticity. For our galaxy, the gradient of DM number density $|\text{d} \ln n_{\phi}/\text{d} r| \sim 1/a$  with Isothermal profile is  orders of magnitude smaller than that inside the typical AGN sources.  
  So only the local density of DM, $\rho=0.4$~GeV/cm$^3$, makes an entry in the first term in eq.~\eqref{pc}, otherwise, it is not sensitive to the local halo profile.

So far, as reported, only one source could be traced back from the observation of astrophysical neutrino at IceCube~\cite{Aartsen:2018ywr}. 
  With more statistics from the next generation of IceCube and other neutrino telescopes like KM3NeT, it might become possible to point out more such sources of astrophysical neutrinos. 
Exploiting the theoretical relations between $\mbh$, $\mh$, and $\msol$, for a subset of these sources  a dedicated fit of the track to shower ratios at various energy bins will provide sensible values of $G_F'$ and $\mdm$.
These can in turn be used to probe other astrophysical neutrino sources. 
With a significant livetime of the future neutrino telescopes we can hope to explore such interesting aspects of neutrino astronomy. This proposed method may then complement other modes of astronomy, in shedding light on the inner dynamics of astrophysical objects.

\vspace{20pt}

\noindent \textbf{Acknowledgments}

\noindent SR thanks Amitava Raychaudhuri and Debanjan Bose for discussions. 
The present work is supported by the Department of Science and
Technology, India {\it via} SERB grants MTR/2019/000997 and CRG/2019/002354.

\end{document}